\documentclass[11pt,a4paper]{article}

\usepackage[a4paper,text={16.3cm,22cm}]{geometry}
\usepackage{amsmath,amsfonts,mathtools,braket,slashed,amssymb,bm,bbm,graphicx}
\usepackage[table,dvipsnames]{xcolor}
    \definecolor{Lightgray}{gray}{0.93}
\usepackage[labelfont=bf,width=0.95\textwidth]{caption}
\usepackage{jheppub}
\hypersetup{
    bookmarksnumbered=true,
    urlcolor=blue,
    linkbordercolor=red,
    citebordercolor=green,
    bookmarksopen=true
}

\allowdisplaybreaks

\numberwithin{equation}{section}

\newcommand{\spac}{{\hspace{0.3mm}}}

\newcommand{\OXaff}{Rudolf Peierls Centre for Theoretical Physics, Department of Physics, Parks Road\\ University of Oxford, Oxford OX1 3PU, UK}
\newcommand{\MITaff}{MIT Center For Theoretical Physics – A Leinweber Institute, Cambridge, MA 02139, USA}
\newcommand{\AECaff}{Albert Einstein Center for Fundamental Physics, Institut für Theoretische Physik\\ Universität Bern, Sidlerstrasse 5, CH-3012 Bern, Switzerland}
\newcommand{\CORNELLaff}{Department of Physics \& LEPP, Cornell University, Ithaca, NY 14853, USA}
\newcommand{\MAINZaff}{PRISMA$^{++}$ Cluster of Excellence \& Mainz Institute for Theoretical Physics\\ Johannes Gutenberg University, Staudingerweg 9, D-55128 Mainz, Germany}
\newcommand{\CERNaff}{CERN, Theoretical Physics Department, CH-1211 Geneva 23, Switzerland}

\title{Non-Global and Super-Leading Logarithms in Higgs Production through Vector Boson Fusion}

\author[a]{Thomas Becher,}
\author[b]{Patrick Hager,}
\author[c,d]{Matthias Neubert,}
\author[e]{Nicolas Schalch,}
\author[c]{Josua Scholze,}
\author[f]{Dominik Schwienbacher}

\affiliation[a]{\AECaff}
\affiliation[b]{\CERNaff}
\affiliation[c]{\MAINZaff}
\affiliation[d]{\CORNELLaff}
\affiliation[e]{\OXaff}
\affiliation[f]{\MITaff}

\preprint{{\small MITP-26-056, CERN-TH-2026-277, OUTP-26-13P, MIT-CTP/6140}}

\abstract{
We present the first combined phenomenological study of non-global and super-leading logarithms in vector-boson-fusion Higgs production and its QCD-induced Higgs-plus-dijet background. A veto on radiation between the widely separated tagging jets provides a natural setting to investigate correlated soft emission and Glauber-induced violations of color coherence. We resum the leading non-global logarithms with the leading-color parton shower framework~\textsc{Marzili} and the super-leading series with Glauber exchanges at $N_c=3$. Combining the two resummed contributions, we study their impact on gap cross sections and on the relative suppression of signal and background.
}

\notoc  

\begin{document}
\maketitle

\clearpage
\section{Introduction}\label{sec:introduction}
Understanding the Higgs sector is central to testing the mechanism of electroweak symmetry breaking and searching for physics beyond the Standard Model. Following the discovery of the Higgs boson~\cite{ATLAS:2012yve,CMS:2012qbp}, measurements of its production and decay modes provide complementary probes of its interactions. Vector-boson fusion (VBF) is particularly useful in this program: the Higgs boson couples directly to the electroweak gauge bosons radiated from two scattering quarks, while the accompanying jets provide a distinctive experimental signature. Selecting two tagging jets with a large invariant mass and a large rapidity separation enhances this contribution relative to other sources of Higgs-plus-dijet events.

The dominant background to VBF arises from QCD-induced Higgs production in association with two jets. The different color connections in the two processes are reflected in their radiation patterns. In the dominant VBF configurations, color flows along each quark line, with no color exchange between the two lines at Born level. Radiation into the central region is therefore suppressed relative to QCD-induced Higgs production. Imposing a veto on radiation in the rapidity gap between the two leading jets provides an efficient way to reduce the background, an idea proposed long ago in~\cite{Dokshitzer:1991he,Bjorken:1992er}. It was studied in detail for the LHC in~\cite{Barger:1994zq,Rainwater:1996ud,Buckley:2021gfw} and has been adopted in ATLAS~\cite{ATLAS:2014aga, ATLAS:2022ooq} and CMS~\cite{CMS:2013zmy,CMS:2014wdm} measurements, which veto jets with transverse momenta above $Q_0=20-30\,\mathrm{GeV}$ between the leading jets. Jet vetoes are also applied in other VBF processes, e.g.\ in $Z$-boson production in association with two jets \cite{CMS:2017dmo,ATLAS:2020nzk}. The paper \cite{CMS:2017dmo} found that the dependence on the veto scale $Q_0$ is not well modeled by existing parton showers.

Fixed-order computations of VBF have reached a very high level of precision. A major milestone was the first fully differential next-to-next-to-leading-order (NNLO) QCD calculation~\cite{Cacciari:2015jma}, which made it possible to predict the Higgs and tagging-jet kinematics at this accuracy. It was performed within the structure-function approximation~\cite{Han:1992hr}, which views the process as two separate scatterings of the incoming quarks and neglects corrections from gluon exchanges between the two quark lines. The NNLO calculation showed that corrections of about one percent to the inclusive rate can become several percent after VBF cuts, demonstrating the importance of controlling radiation for fiducial predictions. Fully differential NNLO predictions are also available in the antenna-subtraction framework~\cite{Cruz-Martinez:2018rod} as well as in the nested soft-collinear subtraction scheme~\cite{Asteriadis:2021gpd}. Remarkably, differential $\rm N^3LO$ predictions in the structure-function approach are known as well~\cite{Dreyer:2016oyx,Dreyer:2018qbw}. Corrections beyond the structure-function approximation were considered in \cite{Liu:2019tuy,Asteriadis:2023nyl,Long:2023mvc}. Such non-factorizable corrections first arise at NNLO and are color-suppressed but enhanced by a factor $\pi^2$, similar to the super-leading logarithms~\cite{Forshaw:2006fk} discussed below. The pure fixed-order results are complemented by NLO calculations matched to parton showers. For VBF, such predictions are available in \textsc{POWHEG}~\cite{Nason:2009ai} and \textsc{aMC@NLO}~\cite{Frixione:2013mta}, the latter within the \textsc{MadGraph5\_aMC@NLO} framework~\cite{Alwall:2014hca}. Comparisons of fixed-order and shower-matched calculations have examined jet definitions, veto requirements, and the remaining shower and matching uncertainties~\cite{Buckley:2021gfw,Barone:2025jey}. The VBF production rate measured in the recent ATLAS combination is compatible with the Standard Model prediction within present uncertainties~\cite{ATLAS:2026pdi}.

The VBF fiducial cuts require the two leading jets to be well separated in rapidity and in opposite hemispheres, favoring configurations in which the hardest jet activity lies away from the center~\cite{Cacciari:2015jma,Buckley:2021gfw}. 
In the region between the two leading jets, one vetoes jets with transverse momentum above a scale $Q_0$, while radiation outside this interval is unrestricted. This realizes the canonical gaps-between-jets geometry: a restriction on part of the angular phase space, with a veto scale that can be much smaller than the characteristic hard scale~$Q$. The resulting scale hierarchy generates enhanced powers of $L=\ln(Q/Q_0)$, whose resummation is essential for a reliable description of the gap cross section and the subject of the present paper.

The angular restriction introduces correlations between radiation inside and outside the gap. A gluon outside the gap is allowed by the measurement and can radiate a softer gluon into it. If the latter emission exceeds the veto scale, its real contribution is rejected while the virtual contribution remains, preventing an inclusive real--virtual cancellation. This mechanism generates non-global logarithms (NGLs), starting at relative order $\alpha_s^2L^2$~\cite{Dasgupta:2001sh,Dasgupta:2002bw}, and was not accounted for in an early soft-gluon resummation study of Higgs-plus-dijet production~\cite{Forshaw:2007vb}. The wide-angle soft radiation generates a single-logarithmic series, $\alpha_s^nL^n$, to which NGLs contribute already at leading-logarithmic (LL) accuracy. Their resummation must therefore account for the new sources of radiation created by successive soft emissions. In the large-$N_c$ limit, this evolution is described by a nonlinear equation~\cite{Banfi:2002hw}, or equivalently by a shower of radiating color dipoles~\cite{Dasgupta:2001sh,Balsiger:2018ezi}. A leading-color treatment of soft radiation also underlies most practical parton showers. The \textsc{PanScales} showers~\cite{Dasgupta:2020fwr,vanBeekveld:2022zhl,vanBeekveld:2022ukn} include classes of subleading-color contributions systematically~\cite{Hamilton:2020rcu}, and their VBF variant is NLL accurate for global observables and resums NGLs at LL accuracy and leading color~\cite{vanBeekveld:2023chs}. A full-color resummation of NGLs is very challenging. First results for the emission from a color dipole were obtained in \cite{Hatta:2013iba}, based on stochastic Wilson-line evolution~\cite{Weigert:2003mm}. More recently, this approach has also been applied to Higgs-plus-dijet partonic configurations~\cite{Hatta:2020wre}. Surprisingly, in the examples considered in this reference, the strict large-$N_c$ limit agreed almost perfectly with the $N_c=3$ results, while the agreement was much worse for various improvement prescriptions proposed in the literature. However, in the formalism used in~\cite{Hatta:2020wre}, Glauber phases cannot be accounted for. This is an important limitation since these phases are responsible for the super-leading logarithms (SLLs) discussed below. Alternatively, one can perform finite-$N_c$ computations in a parton-shower framework, but this requires evolving scattering amplitudes in color space~\cite{Nagy:2007ty,Platzer:2012np,AngelesMartinez:2018cfz,Forshaw:2019ver}, as implemented in the \textsc{CVolver}~\cite{DeAngelis:2020rvq} and \textsc{Deductor}~\cite{Nagy:2019pjp} codes. Within the amplitude-evolution framework, studies of subleading-color corrections to NGLs in four-parton configurations are now available~\cite{Forshaw:2025bmo,Forshaw:2025fif}, together with exploratory steps to include Glauber phases.

Over the past few years, next-to-leading-logarithmic (NLL) resummation of NGLs at leading color has been achieved in three frameworks: the \textsc{Gnole} approach based on nonlinear evolution equations~\cite{Banfi:2021owj,Banfi:2021xzn}, the \textsc{Marzili} framework~\cite{Becher:2021urs,Becher:2023vrh} based on effective field theory, and the \textsc{PanScales} parton showers~\cite{FerrarioRavasio:2023kyg}. So far, however, NLL resummations of NGLs at hadron colliders are available only for Drell--Yan \cite{Becher:2023vrh} and Higgs production~\cite{Becher:2026iim} with a central veto. An application to VBF requires the NLO hard functions for four-parton configurations as well as the resummation of subleading jet-clustering logarithms. The two-loop anomalous dimension accounting for clustering effects was recently extracted~\cite{Becher:2026zon}.

At a hadron collider, Glauber exchange between the incoming partons introduces an additional source of logarithms. The associated imaginary phases obstruct the cancellation of initial-state collinear radiation and generate SLLs~\cite{Forshaw:2006fk,Forshaw:2008cq}. These contributions are intrinsically subleading in color and vanish in the strict large-$N_c$ limit at fixed $N_c\alpha_s$. A leading-color shower therefore misses them even if it correctly resums the leading-color NGLs. Although the first SLL appears only at four-loop order relative to the Born process, as $\alpha_s^4L^5$, it initiates a double-logarithmic tower in which each additional power of $\alpha_s$ is accompanied by two powers of $L$. The logarithmic enhancement and the factors of $\pi$ from the Glauber phases may overpower the suppression by color and powers of the coupling, making an assessment of their numerical impact essential. Schematically, the contribution with two Glauber insertions has the fixed-coupling structure
\begin{equation}\label{eq:SLLgeneric}
   \hat\sigma_{2\to M}^{\rm SLL}(Q_0)
   = \hat\sigma_{2\to M}^{\rm Born}\,
   \frac{\alpha_s\spac L}{\pi\spac N_c}
   \left(\frac{N_c\spac\alpha_s}{\pi}\,i\pi L\right)^2
   \sum_{n=0}^\infty c_n
   \left(\frac{N_c\spac\alpha_s}{\pi}\,L^2\right)^n .
\end{equation}
Here, the coefficients $c_n$ encode the hard process and gap geometry; a colorless particle such as the Higgs may accompany the colored final state. The two Glauber phases supply a factor of $\pi^2$, while the prefactor displays the color suppression relative to leading-color soft evolution. Resumming this series requires keeping track of the ordering of collinear emissions and Glauber exchanges, since their color operators do not commute. The all-order resummation of the SLL series using renormalization-group methods in effective field theory was performed in~\cite{Becher:2021zkk,Becher:2023mtx} and subsequent developments extended the resummation to multiple Glauber exchanges~\cite{Boer:2023jsy,Boer:2023ljq}, incorporated running-coupling effects~\cite{Boer:2024hzh}, and generalized it to massive final states~\cite{Banerjee:2025kkq}. At the leading non-zero order in the color expansion, the series with arbitrarily many Glauber exchanges can be resummed in closed form for all incoming channels~\cite{Boer:2024xzy}. An extension of \textsc{Marzili} achieved the first Monte Carlo resummation of the leading non-zero color contribution with two Glauber exchanges~\cite{Schalch:2026xxk}.

In this work, we investigate the combined numerical impact of NGLs and SLLs on a gap veto in Higgs-plus-dijet production. This is the first phenomenological study combining these resummed contributions for both the VBF signal and its QCD-induced background. Within our effective-theory framework~\cite{Becher:2015hka,Becher:2016mmh}, we evaluate LL non-global evolution with the leading-color \textsc{Marzili} framework and calculate the resummed single- and double-Glauber corrections from the color matrices of the Born-level amplitudes. We combine these separately resummed ingredients to explore their joint effect on the relative signal and background rates. A limitation of our result is that we include NGLs only at leading color, so that we lack a systematic treatment of mixed NGL--SLL terms.

A further limitation of our computation is that an additional class of logarithms is not resummed. The large rapidity separation of the leading jets induces high-energy logarithms of the form $\ln(\hat s/\hat{t}\spac)$, where $\hat{s}$ is the partonic center-of-mass energy squared and $\hat{t}$ is the momentum-transfer invariant, relevant to the QCD-induced background. In our factorization framework, this implies the presence of a large scale hierarchy in the partonic hard functions, with associated large logarithms. The \textsc{High Energy Jets} (HEJ) framework resums these logarithms, and studies of Higgs-plus-dijet production find substantial effects on the background selected by VBF cuts~\cite{Andersen:2017kfc,Andersen:2018kjg}. To achieve full control over the jet-veto cross section, one should perform a combined resummation of all enhanced terms. 

Finally, we only consider the QCD-induced Higgs-plus-dijet background. Depending on the Higgs decay channel, backgrounds without a Higgs boson, such as top-quark pair or diboson production~\cite{ATLAS:2022ooq}, are also relevant and would require separate calculations.

\section{Methodology}\label{sec:methodology}

\subsection{Resummation framework}\label{sec:eft}

We begin with the introduction of the factorization and RG framework developed in~\cite{Becher:2021zkk,Becher:2023mtx,Boer:2024hzh} and recently applied to $pp\to jj$ in~\cite{Becher:2024nqc}.
At leading power in $Q_0/Q$ using a Born-level hard function, the resummed gap cross section takes the form
\begin{equation}\label{eq:evolved_factorization}
   \sigma_{Hjj}(Q_0)
   =\int\!d\xi_1\int\!d\xi_2\,
     \sum_{l=4}^{\infty}\sum_{m=l}^{\infty}
     \big\langle\bm{\mathcal H}_l(\mu_h)
       \ast\bm U_{lm}(\mu_h,\mu_s)
       \otimes\bm{\mathcal W}_m(\mu_s)\big\rangle .
\end{equation}
As the Higgs is color neutral, the Born subprocess $ab\to Hcd$ can be constrained to four colored legs $l=4$.
The hard function $\bm{\mathcal{H}}_4$ contains the Born amplitudes, phase space, and fiducial cuts, and is an operator in color space.
The operator $\bm{U}$ denotes the path-ordered evolution generated by the hard anomalous dimension $\bm{\Gamma}^H$ from $\mu_h\sim Q$ to $\mu_s\sim Q_0$
\begin{equation}\label{eq:evolution_operator}
   \bm U(\mu_h,\mu_s)
   =\mathbf P\exp\!\left[
       \int_{\mu_s}^{\mu_h}\frac{d\mu}{\mu}\,
       \bm\Gamma^H(\mu)
     \right],
   \qquad \bm U(\mu_h,\mu_h)=\bm 1 \, .
\end{equation}
It generates arbitrarily many additional colored directions, so retaining a Born hard process does not truncate the multiplicities generated by the evolution.
The angular brackets denote a color trace, while $\ast$ and $\otimes$ include momentum-fraction and angular convolutions.

The hard anomalous dimension can be expressed as
\begin{equation}\label{eq:anomalous_dimension_split}
\begin{aligned}
   \bm\Gamma^H(z_1,z_2,\mu)
   ={}&\delta(1-z_1)\delta(1-z_2)\,\bm\Gamma^S(\mu)\\
      &+\bm\Gamma^C_1(z_1,\mu)\,\delta(1-z_2)
       +\delta(1-z_1)\,\bm\Gamma^C_2(z_2,\mu)\,,
\end{aligned}
\end{equation}
where $z_i$ are the momentum fractions in the Mellin convolutions contained in $\ast$; they are distinct from the incoming fractions $\xi_i$. The terms $\bm\Gamma_i^C$ describe collinear splittings that redistribute the incoming momentum and can change the parton flavor.

The soft part of the anomalous dimension can be written as
\begin{equation}\label{eq:anomalous_dimension_soft_part}
   \bm\Gamma^S(\mu)
   =\frac{\alpha_s(\mu)}{4\pi}
      \left[\overline{\bm\Gamma}+\bm V^G
            +\bm\Gamma^c\ln\frac{\mu^2}{\mu_h^2}\right]
      +\mathcal O(\alpha_s^2)\, .
\end{equation}
Here, $\overline{\bm{\Gamma}}$ describes wide-angle soft radiation subject to the gap veto, $\bm{\Gamma}^c$ describes initial-state soft-collinear radiation, and $\bm{V}^G$ contains the Glauber phases.
We absorb the one-loop cusp coefficients into $\bm{\Gamma}^c$ and $\bm{V}^G$.
Explicit operator definitions are given in~\cite{Becher:2021zkk,Becher:2023mtx,Boer:2024hzh}.

We use the tree-level low-energy matrix element
$\bm{\mathcal{W}}_m(\mu_s) = f_1(\xi_1,\mu_s) f_2(\xi_2,\mu_s)\bm{1} + \mathcal{O}(\alpha_s)$, where $f_i$ is the parton distribution function (PDF) for incoming parton $i$.
In the numerical analysis, the PDFs are evaluated at the hard factorization scale $\mu_h$ used in the Born-event weights to approximate their scale evolution.
Purely collinear evolution effects beyond this prescription are subleading for the leading double-logarithmic SLL series and are not included.

For the prediction of NGLs, we omit Glauber insertions, so that the soft-collinear terms cancel. We evaluate the remaining evolution generated by $\overline{\bm{\Gamma}}$ at leading color using $\textsc{Marzili}$~\cite{Becher:2023vrh}, as described in Section~\ref{sec:ngl}.
This allows for a resummation of the single-logarithmic series $\alpha_s^n L^n$.

For the SLL correction, we retain up to two insertions of $\bm{V}^G$, one insertion of $\overline{\bm{\Gamma}}$, and arbitrarily many insertions of $\bm{\Gamma}^c$, evaluated at finite $N_c=3$.
The non-commutativity of the Glauber and soft-collinear operators generates terms $\alpha_s^{n+3}L^{2n+3}$ at fixed coupling, where $n$ counts the number of cusp insertions.
We resum this contribution using the running-coupling evolution of~\cite{Boer:2024hzh}; the implementation details are given in Section~\ref{sec:sll}.
The two contributions are evaluated separately and combined using the prescriptions in Section~\ref{sec:results} with
the Born-event samples, fiducial cuts, and hard-scale choices specified in Section~\ref{sec:born_events}.

\subsection{Details on the NGL resummation}\label{sec:ngl}
The starting point is the soft operator $\overline{\bm\Gamma}$ in Section~\ref{sec:eft}, with Glauber insertions omitted. We use \textsc{Marzili}, the Monte Carlo framework introduced for NLL non-global resummation in~\cite{Becher:2023vrh}, restricting the calculation here to its LL evolution. In the leading-color limit, interference between distinct planar color flows is suppressed, and only partons adjacent along a color line form radiating dipoles. For a specified flow, the color operators therefore reduce to numerical factors multiplying dipole kernels. Real emissions still change the set of dipoles, so the evolution continues to couple arbitrarily high multiplicities.

\begin{figure}[p]
\centering
\includegraphics{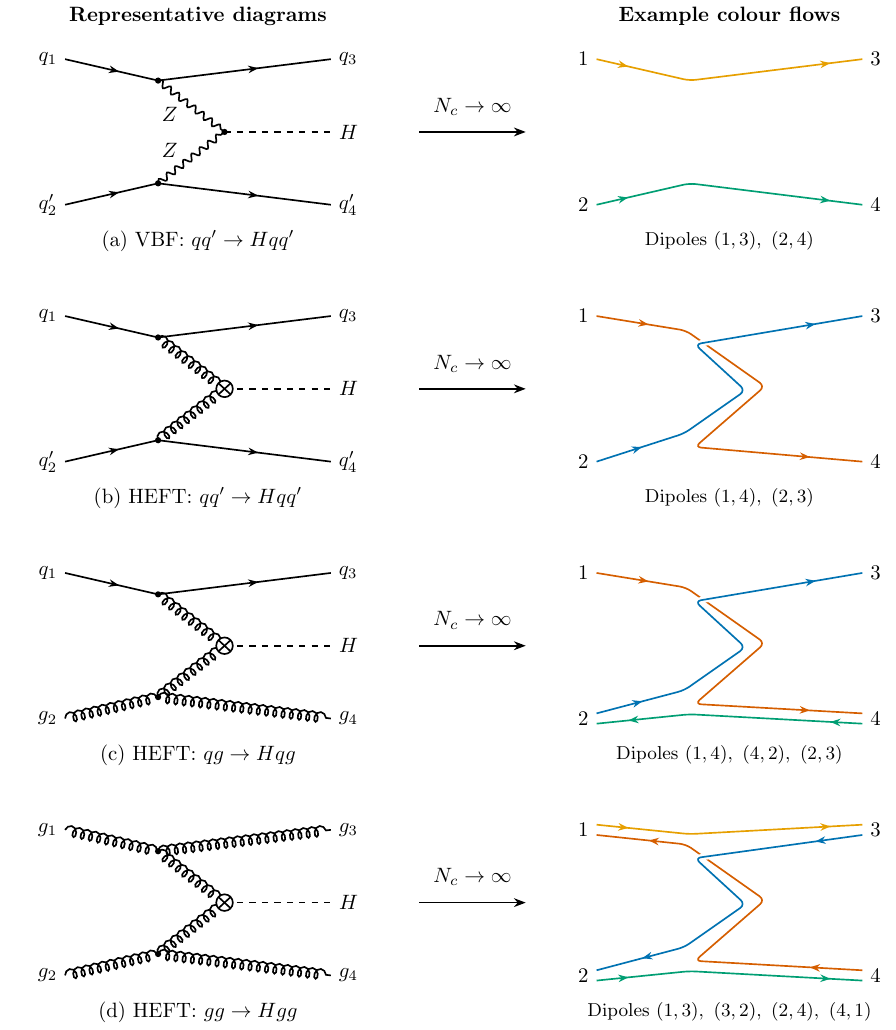}
\caption{Representative hard diagrams (left) and example leading-color flows (right) for the VBF signal and three QCD-induced background channels. The crossed circle denotes the local Higgs-gluon vertex in HEFT. Incoming legs are $1,2$ and outgoing legs are $3,4$; the same external-leg geometry is used in both columns. A gluon carries two oppositely directed color lines. Each continuous color line joining external legs defines a shower dipole, as listed below each flow; crossings are not interaction vertices. For channels with gluons, one possible flow is shown. The shower uses the flow assigned to each event.}
\label{fig:color_flows}
\end{figure}

In this representation, each dipole joins a fundamental color index on one leg to an antifundamental index on another ($\mathbf 3$ and $\overline{\mathbf 3}$ for $N_c=3$). Following the color-space conventions of Catani and Seymour~\cite{Catani:1996vz}, an incoming quark is treated as an outgoing antiquark for the color algebra, and vice versa. Thus the incoming-outgoing quark dipole $(1,3)$ pairs the antifundamental index of incoming leg $1$ with the fundamental index of outgoing leg $3$, see Figure \ref{fig:color_flows}. The directions $n_i$ still follow the physical particle momenta. The labels $(i,j)$ identify the two legs carrying the connected indices. Since a gluon carries both a color and an anticolor index, it is shared between two dipoles.

We absorb the running coupling into the evolution time
\begin{equation}\label{eq:shower_time}
   t(\mu_h,\mu_s)
   =\int_{\mu_s}^{\mu_h}\frac{d\mu}{\mu}\,
      \frac{\alpha_s(\mu)}{4\pi}
   =\frac{1}{2\beta_0}
      \ln\frac{\alpha_s(\mu_s)}{\alpha_s(\mu_h)}\, ,
\end{equation}
with one-loop running, $\beta_0=(11N_c-2n_f)/3$, $n_f=5$, and $\alpha_s(m_Z)=0.118$, so that $t=0$ at the hard scale $\mu_h$ and $t$ increases towards the soft scale $\mu_s$. At leading color, each line connecting two legs $i$ and $j$ forms a dipole $(i,j)$~\cite{Dasgupta:2001sh,Balsiger:2018ezi,Becher:2023vrh}. A dipole emits a soft gluon in the direction $n_k$ with the real and virtual terms
\begin{equation}\label{eq:dipole_generator}
   R_{ij}(n_k)=4N_c\,W_{ij}^k\,\Theta_{\rm out}(n_k)\,,
   \;\;
   V_{ij}=-4N_c\!\int\!\frac{d\Omega_k}{4\pi}\,W_{ij}^k\,,
   \;\;
   W_{ij}^k=\frac{n_i\cdot n_j}{(n_i\cdot n_k)(n_j\cdot n_k)}\,,
\end{equation}
where $n_i=(1,\widehat{\boldsymbol n}_i)$, $\Theta_{\rm out}$ equals one outside the gap and zero inside it, and the collinear singularities are regulated by a common cutoff. Between emissions, the virtual terms exponentiate into the no-emission probability $\exp[t\sum_{[ij]}V_{ij}]$, where $\sum_{[ij]}$ runs over all dipoles of the current configuration~\cite{Balsiger:2018ezi,Becher:2023vrh}.

In the Monte Carlo realization, the first emission into the gap that is not clustered into an existing jet terminates the evolution and sets the soft endpoint of that history. At LL, strong ordering makes it the hardest emission in the gap: for a veto below its scale, the configuration already fails the measurement, and subsequent softer emissions cannot restore it. An emission outside the gap splits the emitting dipole: $(i,j) \mapsto (i,k) \ \ (k,j)$, adding the gluon direction $n_k$ as a new source of softer radiation. The updated dipole set determines the subsequent virtual evolution and real emissions. These successive changes of the emitting system generate the NGLs. All existing directions are kept fixed in the soft approximation, and consequently we do not redistribute any recoil.

For Higgs-plus-dijet production, we initialize the non-global evolution with the Born events generated by \textsc{MadGraph5\_aMC@NLO} \cite{Alwall:2014hca}, as specified in Section~\ref{sec:born_events}. The QCD-induced background is evaluated in the heavy-top effective theory (HEFT). Each event supplies the four colored directions and a color flow through its Les Houches color tags. Figure~\ref{fig:color_flows} illustrates representative hard diagrams and associated leading-color flows.

For the VBF example, the two quark lines give independent dipoles $(1,3)$ and $(2,4)$, whereas the QCD-induced quark background has the connections $(1,4)$ and $(2,3)$. The quark-gluon and four-gluon background examples contain three and four dipoles, respectively. These are the initial dipole sets for the shower; each accepted emission splits the emitting dipole. Other channels and color assignments are treated in the same way using the event's color tags, without reconstructing interference between flows. In particular, independent color lines remain separate throughout the shower.

The different color connections underlie the distinct radiation patterns of signal and background~\cite{Forshaw:2007vb}. In typical VBF kinematics, the signal dipoles connect each incoming parton to a nearby forward jet, suppressing radiation into the central gap. The crossed connections in the quark background instead link incoming partons to jets on the opposite side, favoring radiation between the tagging jets. Gluon channels add further radiating dipoles. Since each dipole contributes through its angular kernel in~\eqref{eq:dipole_generator}, both the number of dipoles and their geometry determine the radiation into the gap.

As the cuts in~\eqref{eq:vbf_cuts} require $y_3y_4<0$, we label the tagging jets such that $y_4<0<y_3$. The gap consists of the rapidity interval between the two jets, and we impose the veto on jets within the gap region. For a single emission into the gap, as relevant for the SLL computation described in Section~\ref{sec:sll}, the clustering is trivial and one obtains the same veto region as in~\cite{Becher:2024nqc}, which consists of the rapidity interval with circular cones of radius $R$ around the jet axes removed,
\begin{equation}\label{eq:gap_region}
   \Omega_{\rm gap}
   =\big\{(y,\phi):\ y_4<y<y_3\,,\ \ \Delta R_{3}>R\,,\ \ \Delta R_{4}>R\big\}\,,
   \qquad R=0.4 \,,
\end{equation}
where $\Delta R_j^2=(y-y_j)^2+(\phi-\phi_j)^2$ is the squared distance to the axis of jet $j=3,4$. For NGLs arising from multiple emissions, we perform a strongly ordered jet clustering as described in \cite{Becher:2023znt}. In this case, the NGL shower terminates if an emission falls within the gap and is not clustered into one of the hard jets outside the gap, i.e.\ not within radius $R$ of one of the existing jets. 

We evaluate the leading-color kernels with $N_c=3$. For each Born event $i$, the shower yields a gap-survival factor $F_i(t)$, with $F_i(0)=1$ corresponding to no evolution. We combine this factor with the original LHE weight $B_i$, normalized so that the selected Born cross section is $N^{-1}\sum_i B_i$, where $N$ is the number of selected Born events. The resulting NGL-resummed gap cross section is
\begin{equation}\label{eq:shower_average}
   \sigma^{\rm NGL}(Q_0)
   =\frac{1}{N}\sum_{i=1}^{N} B_i\,
      F_i\!\left(t(\mu_{h,i},\mu_{s,i})\right).
\end{equation}
The conversion from the veto scale to evolution time is performed separately for each event before averaging. The scale profiles used for this conversion are specified in Section~\ref{sec:results}.

\subsection{Details on the SLL resummation}\label{sec:sll}
The renormalization-group-improved resummation of super-leading logarithms can be obtained by using~\cite{Boer:2024hzh}
\begin{equation} \label{eq:partonic_SLL}
  \hat{\sigma}^\mathrm{SLL}(Q,Q_0)
   =\sum_{l}
    \big\langle \bm{\mathcal{H}}(\mu_h)\,\bm{X}^T \big\rangle\, 
    \mathbbm{U}_\mathrm{SLL}^{(l)}(\mu_h,\mu_s)\,\varsigma \,.
\end{equation}
Here, $l$ denotes the number of Glauber insertions, $X$ and $\varsigma$ are vectors defined in~\cite{Boer:2024hzh}, and $\mathbbm{U}_\mathrm{SLL}^{(l)}$ is the evolution matrix, which runs the hard function from the hard scale $Q$ to the soft scale $Q_0$ and allows for a renormalization-improved resummation. For pure tree-level QCD processes, it is sufficient to use an even number of Glauber insertions, as these hard amplitudes are real-valued, while odd insertions of Glauber phases can contribute for electroweak processes. In our case, we therefore use single and double Glauber insertions, $l=1,2$, to account for the leading logarithmic contribution. We furthermore omitted the convolution $\ast$, which is only necessary for the first soft emission into the gap, as all other contributions do not enhance the phase space.

In order to perform the actual computations, it proves useful to express both the hard functions and the color matrices in a basis $\{|\mathcal{C}_I\rangle\}$ of color configurations for a given channel. As the Higgs is colorless, the basis can, in principle, be chosen the same way as for $pp\to jj$, which was discussed in~\cite{Becher:2024nqc}. In VBF and its QCD background, however, the number of diagrams and contributions is significantly larger than for $pp\to jj$, which motivates an automated implementation. The pipeline for the calculation of super-leading logarithms is then as follows: Similarly to the starting point of \textsc{Marzili}, we generate the Born events using \textsc{MadGraph5\_aMC@NLO}. For each event, we then employ \textsc{OpenLoops 2}~\cite{Buccioni:2019sur} to extract the color basis as well as the hard functions' representation in the trace basis.

We then use this to determine the color matrices $\mathbbm{U}_\mathrm{SLL}^{(l)}$ and calculate the SLL contribution for each event separately, according to~\eqref{eq:partonic_SLL}, and apply \textsc{ColorMath}~\cite{Sjodahl:2012nk} to simplify the color matrices. We use the same veto region as for the NGLs,~\eqref{eq:gap_region}, but with a single soft emission into the gap. We then obtain a reweighting factor accounting for the SLLs and, to allow for an easy extraction, record it in the original event file. 
This framework automates the SLL calculation and can be extended to other gap-between-jets observables.

While substantial parts of the pipeline were coded by large language models (LLMs), several tests were performed to ascertain its accuracy. First, the result for $pp\to jj$ was completely reproduced using the above pipeline, and the color basis and matrix structures were cross-checked using verified Mathematica code. Secondly, two independent models, Claude's Opus 5.5 and OpenAI's GPT-6 Astra, were used, and their results showed complete agreement for $pp\to jj$, VBF, and its QCD background.  Claude's analysis used the above pipeline relying on \textsc{OpenLoops 2}, while Astra's implementation was carried out using the \textsc{OpenPhysics} framework~\cite{openphysics} with \textsc{Helas}~\cite{Murayama:1992gi} routines generated by \textsc{Aloha}~\cite{deAquino:2011ub} within MadGraph. These checks were performed across a wide range of values for both $Q$ and $Q_0$. For further validation, we checked events against a separate implementation we wrote ourselves and found exact agreement. While LLM-generated code can contain errors, our validation procedure gives us confidence that the final result is indeed correct.

\subsection{MadGraph LO events}\label{sec:born_events}
\label{sec:madgraph}

In this section we give further details on the generation of leading-order events, for which the hard matrix elements are evaluated with \textsc{MadGraph5\_aMC@NLO} version~3.7.0~\cite{Alwall:2014hca}. We consider $13\,\mathrm{TeV}$ proton--proton collisions; the Higgs boson is stable and on shell, with $m_H=125\,\mathrm{GeV}$, and five quark flavors are treated as massless. The generated Les Houches events supply the Born momenta, flavor assignments, color tags, and cross-section weights used in the resummation.

For both the signal and background samples, the multi-particle labels \texttt{p} and \texttt{j} include the gluon and the five light quark flavors and their antiquarks. The signal is generated with the \texttt{sm-no\_b\_mass} model and the process definition is given by
\begin{center}
\verb|p p > h j j $$ w+ w- z QCD=0 QED=3|\,.
\end{center}
Restricting the coupling orders selects electroweak Higgs-plus-dijet production, while the \texttt{\$\$ w+ w- z} syntax removes diagrams with $s$-channel $W$ or $Z$ propagators, following the VBF selection of~\cite{Jager:2020hkz}. It excludes Higgs-strahlung topologies, in which an intermediate vector boson produces a Higgs boson and a second vector boson that decays into the two jets.

The background uses the \texttt{heft-no\_b\_mass} model, in which the top quark is integrated out, and its coupling to the Higgs and gluons is represented by local interactions in the HEFT, with
\begin{center}
\texttt{p p > h j j QED=0}\,.
\end{center}
The condition \texttt{QED=0} selects QCD-induced Higgs-plus-dijet production. Figure~\ref{fig:color_flows} illustrates representative diagrams and flows. Consequently, signal and background are treated separately, and their interference is neglected.

At Born level, the two outgoing QCD partons define the tagging jets. We adopt the anti-$k_T$ jet definition with $R=0.4$~\cite{Cacciari:2008gp} and the VBF selection of~\cite{Cacciari:2015jma},
\begin{equation}\label{eq:vbf_cuts}
\begin{gathered}
   p_{3,T},p_{4,T}>25\,\mathrm{GeV},\qquad
   |y_{3}|,|y_{4}|<4.5\,,\\
   y_{3}y_{4}<0\,,\qquad
   |y_{3}-y_{4}|>4.5\,,\qquad
   m_{jj}>600\,\mathrm{GeV} \,.
\end{gathered}
\end{equation}
For consistency, the same cuts are imposed on both samples.

For the central dynamic-scale event samples, we set $\mu_R=\mu_F=Q_s$ for the signal and $\mu_R=\mu_F=Q_b$ for the full background, where
\begin{equation}\label{eq:dynamic_hard_scales}
   Q_s^2=\frac{m_H}{2}
          \sqrt{\left(\frac{m_H}{2}\right)^2+p_{T,H}^2}\,,
   \qquad
   Q_b=\frac{1}{2}
       \left(\sqrt{m_H^2+p_{T,H}^2}+p_{3,T}+p_{4,T}\right).
\end{equation}
The signal scale follows ref.~\cite{Cacciari:2015jma}; $Q_b$ is half the sum of the final-state transverse masses. As an alternative scale choice for the background, we use $\mu_R=\mu_F=Q_b^{(H_T)}=(p_{T,H}+p_{3,T}+p_{4,T})/2$. This adapts the $\hat H_T/2$ choice used for $Z$-plus-jets in the HEJ study of~\cite{Andersen:2026oge}, replacing the $Z$ transverse momentum by that of the Higgs boson. Only the scale choice is adopted; these events do not include HEJ resummation. Separate fixed-scale samples use $\mu_R=\mu_F=80.4\,\mathrm{GeV}$ for the signal and $600\,\mathrm{GeV}$ for the full background. A hard scale of $600\,\mathrm{GeV}$ corresponds to $m_{jj}$, or equivalently the value of the partonic center-of-mass energy. This is an extreme choice, but as Figure \ref{fig:color_flows} illustrates, the background has not only $t$-channel dipoles, but also emissions from $s$- and $u$-channel configurations, which carry a large scale. To resolve the ambiguities associated with the choice of the hard scale, these contributions should be factorized.

In each sample, this scale also sets the hard scale $Q$ at which the veto evolution starts. Both samples use the NNPDF3.0 NNLO PDF set with $\alpha_s(m_Z)=0.118$~\cite{NNPDF:2014otw} and $n_f=5$, although their hard matrix elements are evaluated at LO. This choice is taken from~\cite{Cacciari:2015jma}, ensuring that our setup matches theirs. With the cuts in~\eqref{eq:vbf_cuts} and the scale $Q_s$, our signal Born cross section agrees with their LO result at the 1\% level and we reproduce the size of their scale band. The PDF order affects the Born weights only beyond our LL accuracy.

\section{Results}\label{sec:results}
\begin{figure}[ht]
\centering
\includegraphics[width=0.85\linewidth]{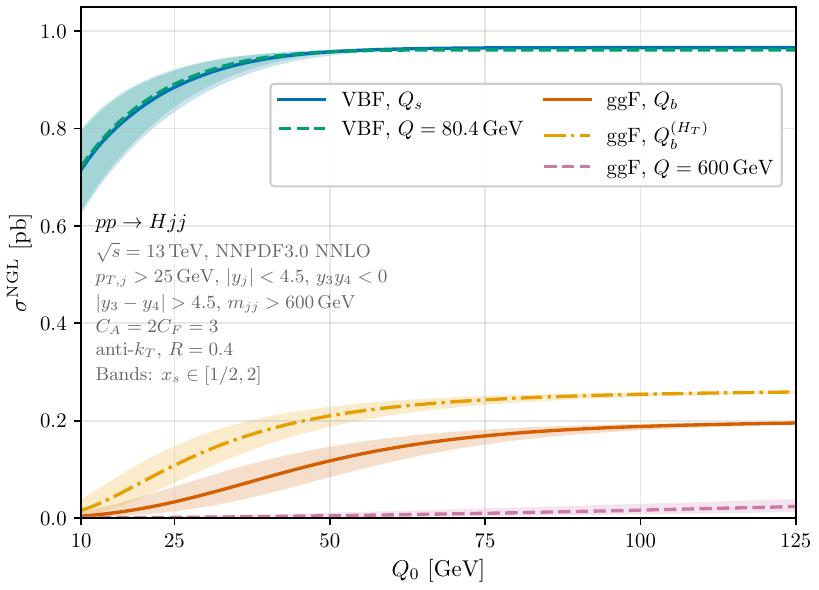}
\caption{Gap cross sections $\sigma^{\rm NGL}$ for the VBF signal and the QCD-induced (ggF) background for the hard-scale choices of Section~\ref{sec:born_events}: $Q_s$ (event scale), $Q_b$ (transverse mass), $Q_b^{(H_T)}$ ($H_T/2$), and the fixed scales $80.4\,\mathrm{GeV}$ and $600\,\mathrm{GeV}$. Lines show $x_s=1$ in~\eqref{eq:profile}, bands the variation $x_s\in[1/2,2]$. Soft emissions are clustered with anti-$k_T$, $R=0.4$.}
\label{fig:NGL_scales}
\end{figure}
\subsection{Non-global logarithms and hard-scale dependence}\label{sec:results_ngl}

The shower yields the gap-survival factors $F_i(t)$ as functions of the evolution time. To obtain the gap cross section as a function of the veto scale, we relate $t$ to $Q_0$ for each event through~\eqref{eq:shower_time}, with $\mu_h=Q$ and a profile function for the soft scale,
\begin{equation}\label{eq:profile}
   \mu_s=\frac{x_s\,Q_0}{\left(1-Q_0/Q\right)^n+x_s\,Q_0/Q}\,,
   \qquad Q_0<Q\,,
\end{equation}
with $n=4$~\cite{Balsiger:2019tne,Becher:2026iim}. The profile reduces to the canonical choice $\mu_s=x_sQ_0$ for $Q_0\ll Q$ and approaches $\mu_s=Q$ for $Q_0\to Q$, so that the resummation is switched off smoothly at the hard scale. For $Q_0\geq Q$, no evolution takes place and $F_i=1$. Once $Q_0$ exceeds the hard scales of all events, $\sigma^{\rm NGL}$ therefore coincides with the LO cross section $\sigma^{\rm Born}$. We use $x_s=1$ as the central value and vary $x_s\in[1/2,2]$.

Figure~\ref{fig:NGL_scales} shows the resulting gap cross sections $\sigma^{\rm NGL}$. The hard scale sets the starting point of the evolution: the larger $Q$, the more running down to a given $Q_0$, and the larger the evolution time $t$. Since the background survival factor decreases rapidly with $t$, this additional evolution suppresses the background strongly, while the signal is hardly affected. Accordingly, the two signal curves nearly coincide, whereas the background gap fraction $\sigma^{\rm NGL}/\sigma^{\rm Born}$ at $Q_0=25\,\mathrm{GeV}$ ranges from $0.41$ for the smallest hard scale, $Q_b^{(H_T)}$, to $0.016$ for the fixed scale of $600\,\mathrm{GeV}$. This spread far exceeds the soft-scale bands, so the hard-scale choice is the dominant uncertainty of the background prediction. NLL resummation~\cite{Becher:2023vrh,Becher:2026iim} would reduce it, but for precise predictions, one should factorize the physics at the different scales in the hard function and resum the associated large logarithms by combining NGL and HEJ resummations. We note that in addition to the strong suppression of the cross section, the choice $600\,\mathrm{GeV}$ induces very large SLLs.

The different behavior of signal and background can be understood from the shower algorithm. Starting from the current configuration, the next emission occurs after an evolution time $\ln r/\sum_{[ij]}|V_{ij}|$, where $r$ is a uniform random number in $(0,1)$ and the virtual terms $V_{ij}<0$ are given in~\eqref{eq:dipole_generator}. Each $|V_{ij}|$ grows with the rapidity range spanned by the dipole, and every additional dipole adds to the sum. If the emission lies in the gap, the evolution of this event terminates; otherwise, the shower continues with the split dipole. In the background, the dipoles span the gap between the tagging jets, and gluon channels supply further dipoles. Emissions therefore occur at smaller $t$, and many of them fall into the gap, so the shower stops at small $t$ much more often than for the signal. In contrast, the dipoles in the VBF signal connect each incoming parton with the jet in the same hemisphere and rarely radiate into the gap. 

All results include anti-$k_T$ clustering of the soft emissions with $R=0.4$~\cite{Cacciari:2008gp}: an emission within a distance $R$ of an existing jet is clustered into it and does not veto the event. Compared with the unclustered results, the gap cross sections change by less than $3\%$ for $Q_0\geq10\,\mathrm{GeV}$, except for the strongly suppressed background evaluated with $\mu_h=600\,\mathrm{GeV}$, where the change reaches $7\%$ at the lowest $Q_0$. The leading clustering effects are therefore small compared to the hard-scale uncertainty.

\begin{figure}
\centering
\includegraphics[width=0.85\linewidth]{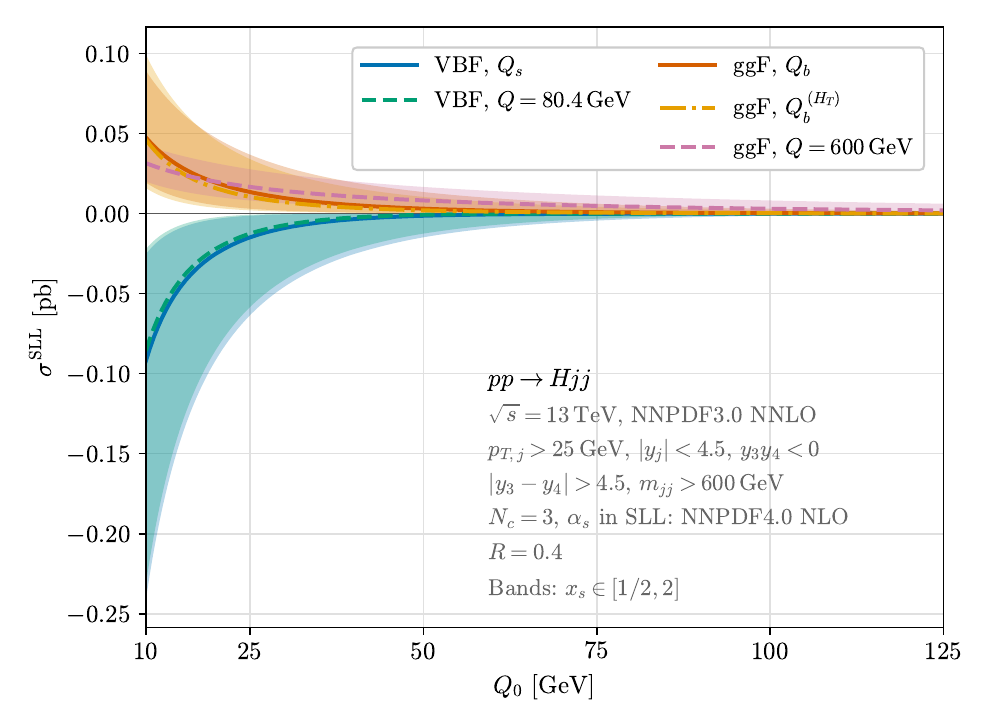}
\caption{The pure SLL contribution for various schemes of the hard scale for the signal and background. The error bands were obtained by scale variation. While the signal is mostly unmodified by super-leading logarithms, except at very low values of $Q_0$, the background receives percent-level corrections. The choice of the different scheme, whether using the $2m_W$, $H_T/2$ or $\hat{H}_T/2$, does not impact the result significantly. While SLLs suppress the signal, the background is enhanced due to the dominance of gluonic channels.}
\label{fig:SLL_all_schemes}
\end{figure}

\subsection{Super-leading logarithms}\label{sec:results_sll}
Figure~\ref{fig:SLL_all_schemes} shows the isolated SLL corrections
$\sigma^{\mathrm{SLL}}$ for VBF (signal) and gluon fusion (background).
The signal correction is negative, whereas the net background correction is positive. As in the NGL case, this difference can be traced back to the partonic channels involved: the signal is dominated by VBF, where fewer color connections are possible, while the background arises mainly from gluonic channels, which yield a correction of the opposite sign.

At $Q_0=25\,\mathrm{GeV}$, the corrections are of similar size for all three hard-scale choices, with $-1.3\%$ for the signal ($\mu_h\sim m_W$) and $+7.5\%$ for the background ($\mu_h \sim 2m_W$), relative to the corresponding Born cross sections.

The differences between the hard-scale choices are smaller than the displayed scale-variation bands, which are obtained by varying the soft scale by a factor of two up and down.
The magnitudes of the corrections decrease as the evolution interval shrinks, and increase as the veto scale $Q_0$ is lowered, ultimately diverging.
The isolated SLL series does not include the suppression from repeated soft emissions, and this divergence should not be interpreted as the behavior of the complete gap cross section for $Q_0\to 0$.
\begin{figure}
\centering
\includegraphics[width=0.85\linewidth]{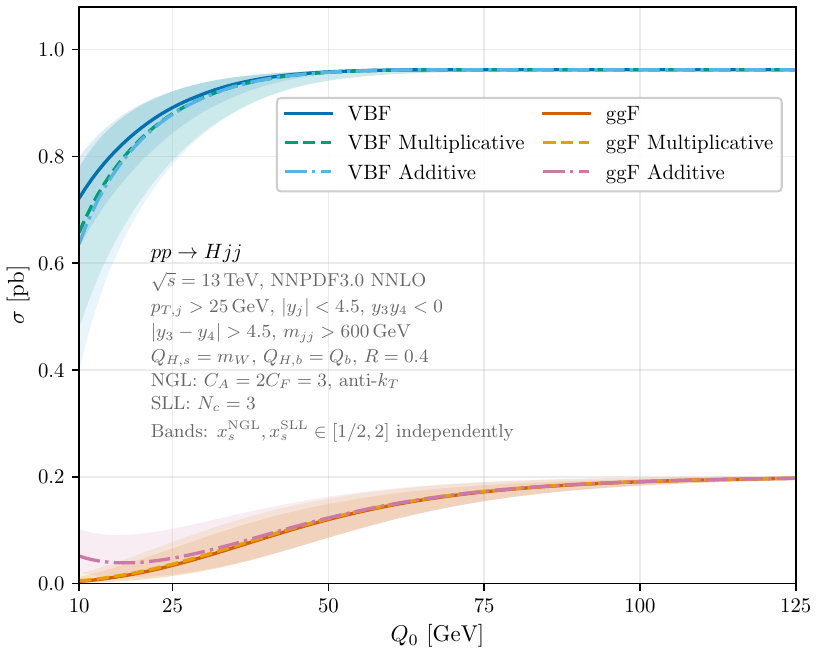}
\caption{The NGL and SLL contributions for different scale choices, as well as different combinations of the two. NGLs are evaluated at large $N_c$, while SLLs use the exact color. }
\label{fig:NGL_and_SLL}
\end{figure}

\subsection{Combination}\label{sec:results_combination}
To obtain a more realistic estimate for the final cross section, we combine the SLL correction with the leading-color NGL prediction via two different methods
\begin{align}
    \sigma_X^{\mathrm{add}} &=\sigma_X^{\mathrm{SLL}}+\sigma_X^{\mathrm{NGL}}\, ,\\
    \sigma_X^{\mathrm{mult}} &=\sigma_X^{\mathrm{NGL}}\frac{\sigma_X^{\mathrm{Born}}+\sigma_X^{\mathrm{SLL}}}{\sigma_X^{\mathrm{Born}}}
    \,,
\end{align}
with $X=S,B$ for signal and background. Note that $\sigma_X^{\mathrm{NGL}}$ includes the Born cross section, while $\sigma_X^{\mathrm{SLL}}$ starts at higher orders. 

The simple additive prescription retains the SLL correction without any soft dressing such that adding both contributions introduces no overlap between the separate sectors, as the NGL baseline omits Glauber exchanges. This approach, however, features the divergent behavior for $Q_0\to 0$ from the SLLs.
The multiplicative prescription instead assumes that the SLL correction experiences the same veto suppression as the Born process, by multiplying it with the NGL resummation.
This is a naive model for the dressing by additional soft radiation corresponding to multiple insertions of $\overline{\bm{\Gamma}}$ in the EFT framework, and we leave the full color-correlated evolution of these insertions for future work. The difference between the two prescriptions consists of products involving additional subleading-color single-logarithmic factors and is beyond the leading double-logarithmic SLL accuracy.
We use the prescription dependence to probe sensitivity to the missing terms.

Figure~\ref{fig:NGL_and_SLL} shows the results for the fixed-scale choice on the VBF signal and $Q_b$ for the background, with the NGL factor evaluated with the central profile in~\eqref{eq:profile}.
The NGL prediction leaves the signal close to its Born value, while strongly suppressing the background.
At $Q_0=25\,\mathrm{GeV}$, the central curves give approximately $0.89\,\mathrm{pb}$ for the signal and $0.01\,\mathrm{pb}\sim 0.126\,\sigma_B^\mathrm{Born}$ for the background.
The SLL contribution partly offsets the background suppression: the additive and multiplicative prescriptions yield background cross sections of about $0.201\,\sigma_B^\mathrm{Born}$ and $0.135\,\sigma_B^{\mathrm{Born}}$, respectively.
Both prescriptions yield larger backgrounds than the NGL-only prediction, but remain well below the Born background. The shaded bands show the envelope of the scale variations.

\section{Conclusion}\label{sec:conclusion}

In this paper, we presented the first combined calculation of non-global and super-leading logarithms for Higgs-boson production via vector-boson fusion, as well as for its QCD background. The non-global soft evolution leaves the signal almost unchanged but strongly suppresses the background, while the SLLs reduce the signal by a few percent and enhance the background by several percent. The first few orders of the NGL result are included in existing fixed-order results and the leading-color NGLs can be resummed using logarithmically accurate parton showers. For the VBF signal, the \textsc{PanScales} collaboration presented such a shower in~\cite{vanBeekveld:2023chs}. The SLLs, on the other hand, are not accounted for in any of the existing predictions. Our results show that subleading-color and coherence-violating effects must be included in future precision studies. 

Our predictions are obtained by connecting a chain of existing computational tools to our effective field theory framework. Starting from a common set of Born events generated with the \textsc{MadGraph5\_aMC@NLO} framework, we compute the NGLs with \textsc{Marzili}, and the SLLs with \textsc{OpenLoops 2}, which provides the color structure of the underlying amplitudes. For each event, both corrections are stored within the event files as functions of the veto scale, so that they can easily be applied when computing differential cross sections. We found AI coding assistants to be very helpful in building such tool pipelines. To verify the results, we performed several cross-checks and compared independent implementations using different AI models against each other and against our earlier code~\cite{Becher:2024nqc}. Our work represents a first step towards the automated calculation of SLLs and NGLs in gaps-between-jets cross sections for a wide class of processes, paving the way for including SLLs in future precision studies.

It remains an open question how the different logarithms should be combined. While NGLs alone certainly do not capture the full picture, it is unclear how the leading SLL contribution should be combined with them. 
The additive prescription inherits the divergence of the SLL result for $Q_0\to 0$, whereas the multiplicative prescription only models the dressing by additional soft radiation. 
Answering this question conclusively will require a joint evolution of NGLs and SLLs beyond leading color, including Glauber phases. The one-loop anomalous dimension for this resummation is available~\cite{Becher:2021zkk}, but solving the associated renormalization-group equation requires evolution in color space~\cite{Nagy:2007ty,Platzer:2012np,AngelesMartinez:2018cfz,Forshaw:2019ver}. First numerical results for subleading-color NGLs for four-parton configurations recently became available~\cite{Forshaw:2025bmo,Forshaw:2025fif}, together with an exploratory study of Glauber phases. Within the effective-theory framework, the leading SLLs for selected partonic configurations have recently been resummed~\cite{Schalch:2026xxk} by extending \textsc{Marzili} beyond leading color, which paves the way for a full treatment of NGL and SLL evolution.

\subsection*{Acknowledgments}

TB thanks J\"urg Haag for discussions of jet-clustering effects. PH thanks Ruben Monten for early access to the \textsc{OpenPhysics} framework and useful discussions on implementing agentic workflows.
NS thanks Alexander Karlberg and Fabrizio Caola for discussions about the scale setting for the VBF signal.
JS thanks Upalaparna Banerjee for discussions.
 
  This research has received funding from the European Research Council (ERC) under the European Union’s Horizon 2022 Research and Innovation Program (ERC Advanced Grant agreement No.~101097780, EFT4jets) and from the Swiss National Science Foundation (SNSF) under grant 200021\_219377. The work reported here was also supported by the Cluster of Excellence \emph{PRISMA${}^{++}$} (EXC 2118/2, Project ID 390831469) funded by the German Research Foundation (DFG) within the German Excellence Strategy. DS was funded under the Swiss National Science Foundation Postdoc.mobility grant Nr. P500-2\_239446. NS is funded through the Royal Society, grant RP\textbackslash R\textbackslash 231001. 

  TB, MN and DS thank the Erwin-Schrödinger International Institute for Mathematics and Physics at the University of Vienna for partial support during the Programme "New Paradigms for Harnessing Quantum Field Theory at Colliders", July 27--August 28, 2026.
  JS thanks the Albert Einstein Center for Fundamental Physics, where part of this work was carried out, for hospitality. 
  
  Views and opinions expressed in this work are those of the authors only and do not necessarily reflect those of the European Union or the European Research Council Executive Agency. Neither the European Union nor the granting authority can be held responsible for them. 

\pdfbookmark[1]{References}{Refs}
\bibliography{refs.bib}

\end{document}